\documentclass[multphys,vecphys]{svmult}

\usepackage{amssymb}

\usepackage{makeidx}         % allows index generation
\usepackage{graphicx}        % standard LaTeX graphics tool
\usepackage{multicol}        % used for the two-column index
\usepackage[bottom]{footmisc}% places footnotes at page bottom
\makeindex             % used for the subject index
\begin{document}

\title*{Lack of cooling flow clusters at $z>0.5$}
% Use \titlerunning{Short Title} for an abbreviated version of
% your contribution title if the original one is too long
\author{A.~Vikhlinin\inst{1,2}\and
R.~Burenin\inst{2} \and
W.~R.~Forman\inst{1} \and
C.~Jones\inst{1} \and
A.~Hornstrup\inst{3} \and
S.~S.~Murray\inst{1} \and
H.~Quintana\inst{4}
}
\authorrunning{Vikhlinin et al.}
% Use \authorrunning{Short Title} for an abbreviated version of
% your contribution title if the original one is too long
\institute{Harvard-Smithsonian Center for Astrophysics, Cambridge MA, USA
\and Space Research Institute, Moscow, Russia
\and Danish National Space Center
\and Dep. de Astronomia y Astrofisica, 
     Pontificia Universidad Catolica de Chile
}
%
% Use the package "url.sty" to avoid
% problems with special characters
% used in your e-mail or web address
%
\maketitle

The goal of this work is to study the incidence rate of ``cooling
flows'' in the high redshift clusters using \emph{Chandra} observations
of $z>0.5$ objects from a new large, X-ray selected
catalog~\cite{2006astro.ph.10739B}. We find that only a very small
fraction of high-$z$ objects have cuspy X-ray brightness profiles, which
is a characteristic feature of the cooling flow clusters at $z\sim0$.
The observed lack of cooling flows is most likely a consequence of a
higher rate of major mergers at $z>0.5$.

\section{Introduction}

The central regions in a large fraction of low-redshift clusters are
clearly affected by radiative cooling \cite{1994ara&a..32..277f}. Some
estimates put the fraction of such cooling flow clusters to $>70\%$
(e.g., \cite{1998MNRAS.298..416P}). A recent by Bauer et al.{} suggests
that the cooling flow fraction remains high to $z\sim
0.4$~\cite{2005MNRAS.359.1481B}. However, this work is based on the
\emph{ROSAT} All-Sky Survey cluster sample, and so it can be strongly
affected by Malmquist bias (strongly over-luminous clusters are
preferentially selected because of the high flux threshold). 

Any evolution in the cooling flow fraction, if detected, must be taken
into account in detailed physical models of this phenomenon. We address
this important question using a new distant cluster sample, derived from
a sensitive survey based on the \emph{ROSAT} pointed observations
\cite{2006astro.ph.10739B}. All objects were observed with
\emph{Chandra}, providing a uniform dataset which should be much less
affected by selection effects than the previous samples.

\section{Definition of the ``cooling flow'' cluster}
\label{sec:oper-defin-cool}

First of all, we need to choose a definition of the ``cooling flow''
cluster that can be efficiently applied to the X-ray data of various
statistical quality. The most common definition is based on the
estimated central cooling time: cooling flow clusters have $t_{\rm
  cool}\ll t_H$ (e.g., \cite{1998MNRAS.298..416P}).  One could also use
the mass deposition rate given by the standard cooling flow model
\cite{1994ara&a..32..277f}; cooling flow clusters have $\dot{M} \gtrsim
(10-100)\,M_\odot$~yr$^{-1}$~\cite{1998MNRAS.298..416P}. These
definitions rely on spatially-resolved spectroscopic measurements which
is a serious disadvantage for application at high redshifts. The data of
sufficient quality to measure $T(r)$ are available only for a small
number of high-$z$ objects. We, therefore, look for a definition based
solely on the X-imaging data.

\begin{figure}[t]
\vspace*{-4.7mm}
\centerline{
  \includegraphics[width=0.5\linewidth]{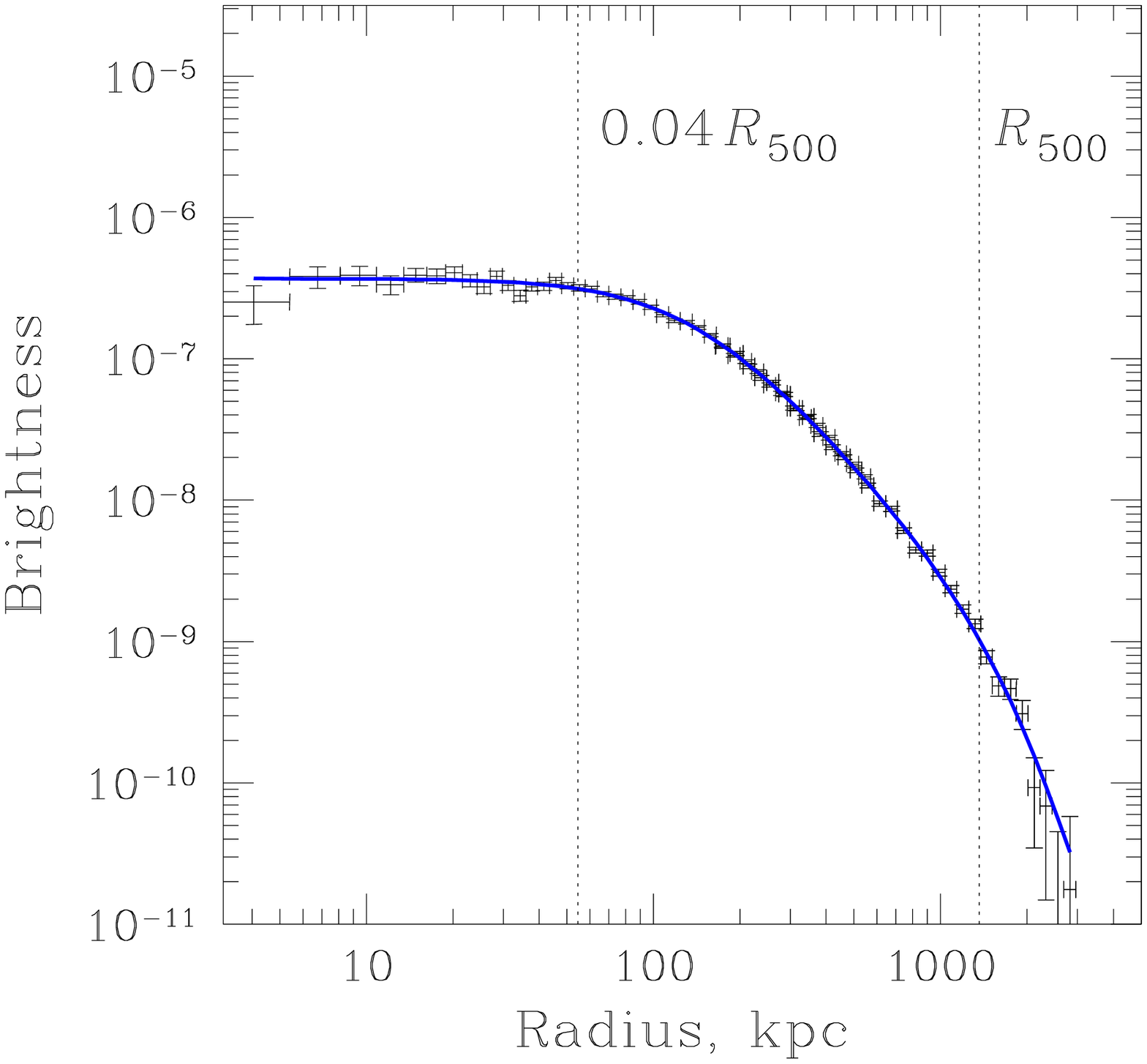}\hfill
  \includegraphics[width=0.5\linewidth]{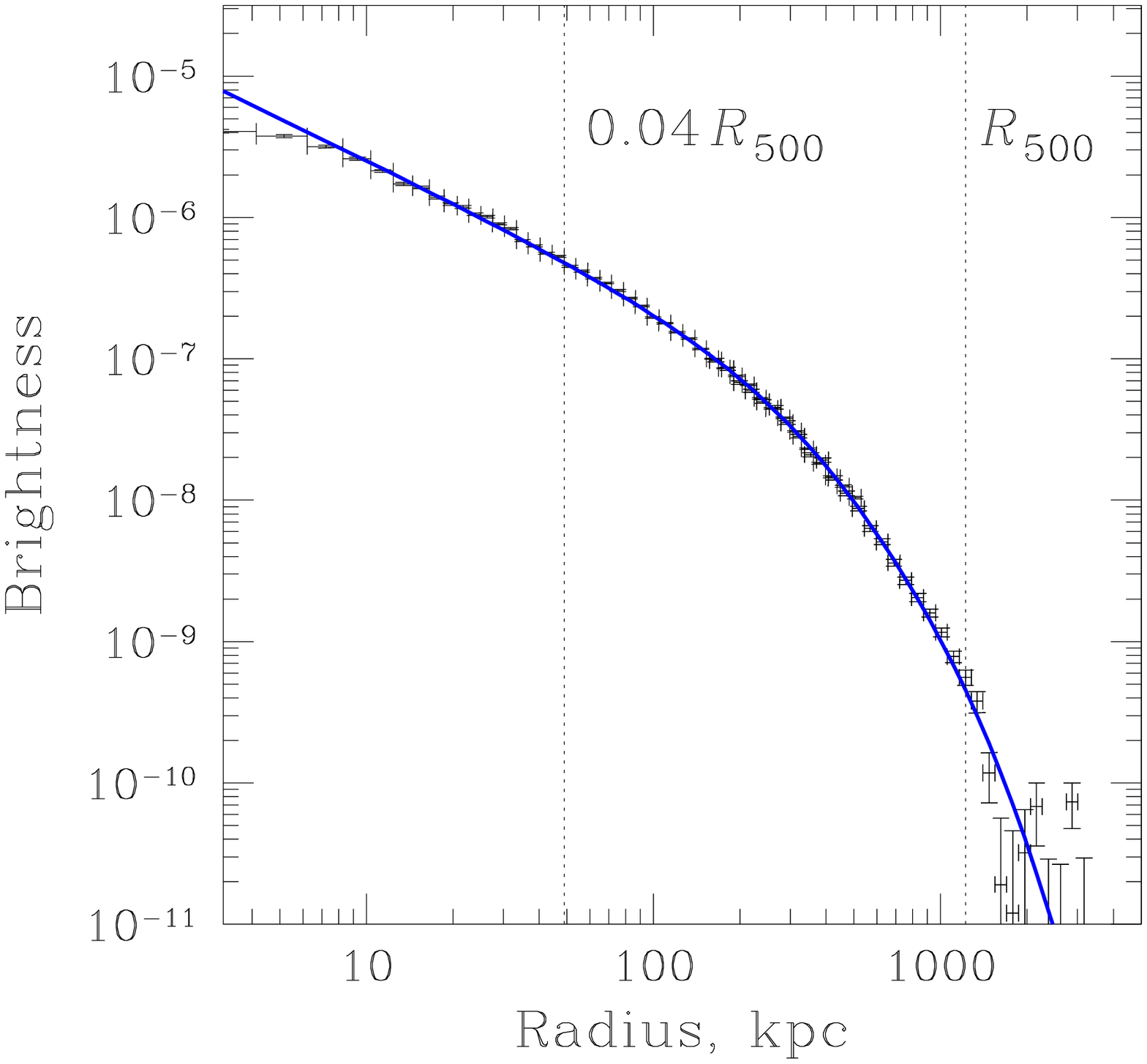}
}
\vspace*{-4.7mm}
\caption{X-ray surface brightness profiles typical for non-cooling flow
  (\emph{left}; A401) and cooling flow (\emph{right}; A85) clusters. 
  Solid lines show the model X-ray brightness corresponding to the
  best-fit gas density model (see \cite{2006ApJ...640..691V} for
  details).} 
\label{fig:prof:examples}
\end{figure}

At low redshifts, there is a clear connection between the presence of
the cooling flow and the X-ray morphology. Clusters with $t_{\rm
  cool}\gtrsim t_H$ have X-ray brightness profiles with flat cores while
those with $t_{\rm cool}\ll t_H$ have characteristic central cusps in
the X-ray brightness distribution (Fig.\,\ref{fig:prof:examples}).  The
central cusp can be characterized by the power-law index of the gas
density profile, $\alpha=d\log \rho_g/d\log r$. For uniformity, the
radius at which $\alpha$ is measured should be chosen at a fixed
fraction of the cluster virial radius. This radius should be sufficiently
small so that the effects of cooling are strong. At very small radii,
however, the density profiles even in clusters with strong cooling flows
can flatten because of the outflows from the central AGN (see many
papers in these proceedings). Empirically, a good choice is
$r=0.04\,R_{500}$,\footnote{$R_{500}$ is the radius at which the mean
  enclosed total mass overdensity is 500 relative to the critical
  density at the object redshift. $R_{500} \approx 0.5 R_{\rm vir}$.}
and so we define the ``cuspiness parameter'', $\alpha$, as
\begin{equation}
\displaystyle\alpha \equiv\frac{d\log \rho_g}{d\log r} \qquad
\mathrm{at} \quad r=0.04\,R_{500}
\end{equation}
Cuspiness can be measured by fitting a realistic 3-dimensional gas
density model to the observed X-ray surface brightness (our procedure is
described in \cite{2006ApJ...640..691V}). Examples of the best-fit
models are shown by solid lines in Fig.\,\ref{fig:prof:examples}.  Such
modeling is feasible with moderate-exposure \emph{Chandra} observations
of high-redshift clusters. $R_{500}$ can be estimated using low-scatter
cluster mass proxies such as the average temperature (excluding the
central cooling region). We use an even better proxy, the recently
proposed $Y_X$ parameter \cite{2006ApJ...650..128K}, which is remarkably
insensitive to the cluster dynamical state and easily measured even in
high-redshift clusters.

\begin{figure}[t]
\vspace*{-4.7mm}
\centerline{
  \includegraphics[width=0.6\linewidth]{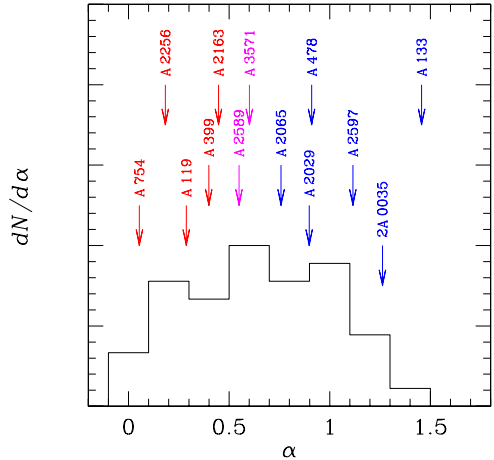}
}
\vspace*{-4.7mm}
\caption{The distribution of the cuspiness parameter in the low-$z$
  cluster sample. Arrows indicate the values for some well-known
  clusters. The boundary value of $\alpha=0.5$ approximately separates
  cooling flow and non-cooling flow clusters.}\label{fig:distr:local}
\end{figure}

Our low-redshift cluster sample is a flux-limited subsample of 48
objects from the HIFLUGCS catalog \cite{2002ApJ...567..716R}, all with
the archival \emph{Chandra} observations. The distribution of the
cuspiness parameter for these objects is shown in
Fig.\,\ref{fig:distr:local}. Clearly, the value of $\alpha$ is closely
connected to the more common cooling flow definitions.  Clusters with
$\alpha>0.7$ (e.g., A2065, A478, A2029, A2597, 2A~0035, A133) are known
to host strong cooling flows.  The objects with $\alpha<0.5$ (e.g.,
A2163, A399, A119, A2256, A754) are famous non cooling flow clusters.
The clusters in the range $0.5<\alpha<0.7$ (e.g., A2589, A3571) host
weak cooling flows.  Therefore, \emph{the cooling flow clusters are
  those with $\alpha>0.5$.}  Approximately 2/3 of the low-redshift
sample (31 of 48 objects) have cuspiness above this value, in line with
the previous estimates of the cooling flow incidence rate
(e.g.~\cite{1998MNRAS.298..416P}).

\section{High-redshift cluster sample}

Our high-redshift sample is derived from the recently completed
400~deg$^2$ \emph{ROSAT} PSPC survey (400d;~\cite{2006astro.ph.10739B}). 
This is the largest-area survey based on the \emph{ROSAT} pointed
observations. Clusters are detected as extended X-ray sources in the
central $17.5'$ of the PSPC FOV and required to have fluxes
$f_x>1.4\times10^{-13}$~erg~s$^{-1}$~cm$^{-2}$. The X-ray sample is
fully identified.  It includes 266 optically confirmed clusters (95\% of
the X-ray candidate list). Spectroscopic redshifts are available for all
objects. 

A subsample of the high-$z$ 400d clusters has been observed with
\emph{Chandra}. The exposure times were chosen to yield at least 2000
source counts, which is sufficient to measure the average cluster
temperature with a $15\%$ uncertainty and accurately trace the surface
brightness profile to $r\sim R_{500}$. \emph{Chandra}'s angular
resolution corresponds to a linear scale of $<8$~kpc out to $z=1$, fully
sufficient to measure the cuspiness parameter. In this work, we use 400d
clusters at $z>0.5$, 20 in total. The typical mass of these objects
corresponds to today's 4~keV clusters.

\begin{figure}[t]
\vspace*{-4.7mm}
\centerline{
  \includegraphics[width=0.5\linewidth]{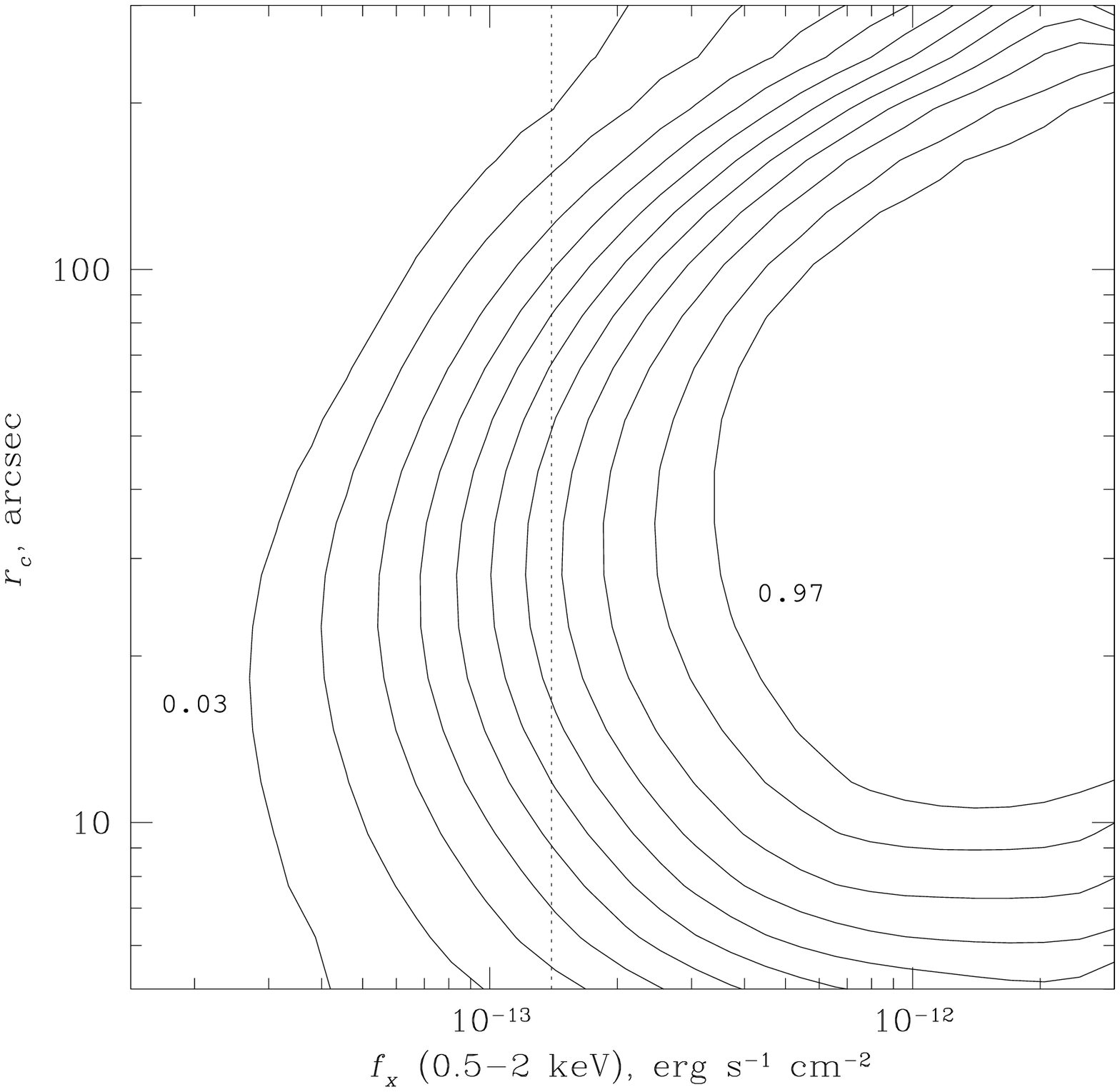}\hfill
  \includegraphics[width=0.5\linewidth]{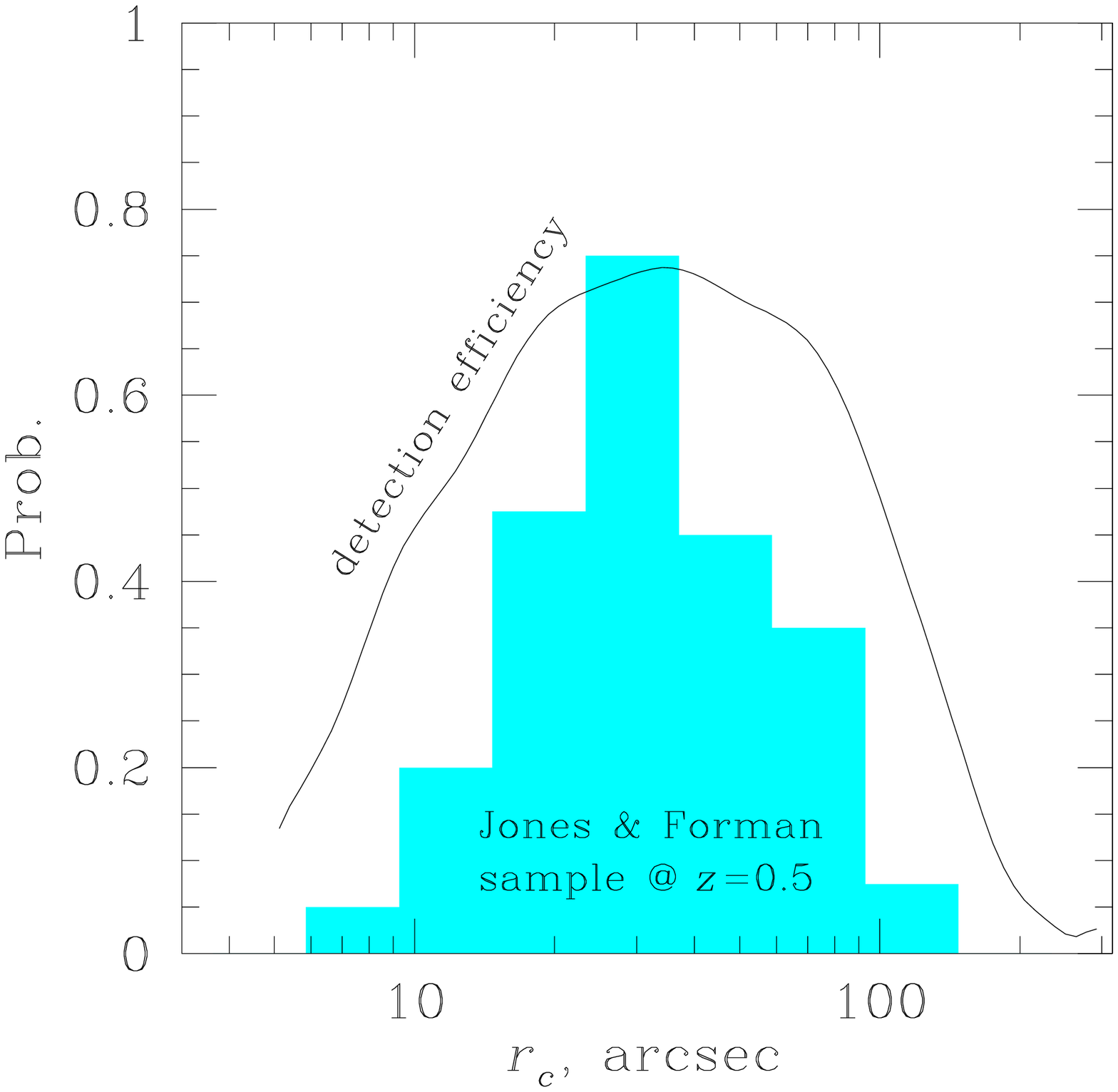}
}
\vspace*{-2.7mm}
\caption{\emph{(a)} --- Detection efficiency of the 400d survey for
  idealized $\beta$-model clusters as a function of total flux and
  core-radius. Dotted line shows the flux limit of the 400d catalog. 
  Detection efficiency is reduced for $r_c\lesssim 8''$ because the such
  clusters are hard do distinguish from the point sources. The
  efficiency is also small for objects with $r_c\gtrsim150''$ because
  they are ``lost'' in the cosmic X-ray background.\quad \emph{(b)}
  Detection efficiency as a function of core-radius for
  $f_x=2\times10^{-13}$~erg~s$^{-1}$~cm$^{-2}$. Shaded histogram shows
  the distribution of core-radii in a low-redshift sample \cite{jf99}
  scaled to $z=0.5$.} 
\label{fig:detprob}
\end{figure}

The basic characteristics of the X-ray selection in the 400d survey have
been extensively calibrated through exhaustive Monte-Carlo simulations
(see~\cite{2006astro.ph.10739B} for details). The aspect most relevant
for the present study is the sensitivity of the detection efficiency to
the cluster size and structure. A precise two-dimensional map of the
detection efficiency as a function of cluster size and core radius was
derived for idealized $\beta$-model clusters (Fig.\,\ref{fig:detprob}a).
The detection efficiency drops significantly only for $r_c\lesssim 8''$
and $r_c\gtrsim 120''$ (see Fig.\,\ref{fig:detprob}b which shows the
slice through the detection probability map at
$f_x=2\times10^{-13}$~erg~s$^{-1}$~cm$^{-2}$, just above the survey flux
limit). The angular size range in which the 400d X-ray detection is
sensitive encompasses the entire range of core-radii expected for the
high-redshift clusters (c.f.~shaded histogram in
Fig.\,\ref{fig:detprob}b).  Therefore, the 400d selection will not bias
the distribution of core-radii for $\beta$-model clusters.

The sensitivity of the 400d X-ray detection algorithm to cooling flow
clusters with the cuspy X-ray brightness profiles requires a separate
study. This issue was addressed by a separate set of Monte-Carlo
simulations in which instead of $\beta$-models, we used the real X-ray
images of a complete sample of low-$z$ clusters, scaled to different
redshifts in the range $0.35<z<0.8$ (see \cite{2006astro.ph.10739B} for
details). A short summary of the results from these simulations is that
there is no significant difference in the detection efficiency for the
$\beta$-model and cooling flow clusters (see, e.g., Fig.~16 in
\cite{2006astro.ph.10739B}). For example, \mbox{Hydra-A} ($\alpha=1.24$)
at $z=0.45$\footnote{Redshifts here are chosen so that the observed
  fluxes would correspond to that in Fig.\,\ref{fig:detprob}b,
  $2\times10^{-13}$~erg~s$^{-1}$~cm$^{-2}$.} is detected with the
probability 0.67; 2A~0335 ($\alpha=1.26$) at $z=0.45$ has $p_{\rm
  det}=0.54$; A2029 ($\alpha=0.90$) at $z=0.8$ has $p_{\rm det}=0.69$.
These values are near the maximum efficiency for $\beta$-model clusters
of similar flux (Fig.\,\ref{fig:detprob}b). Therefore, there should be
no discrimination in the 400d survey against the objects similar to
today's cooling flow clusters.

\section{Observed morphologies and cuspiness parameters of the high-redshift
  clusters} 

\begin{figure}[t]
\centerline{\includegraphics[width=0.4\linewidth]{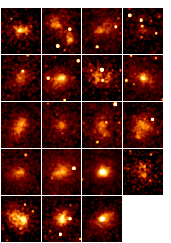}\hfill
  \includegraphics[width=0.6\linewidth]{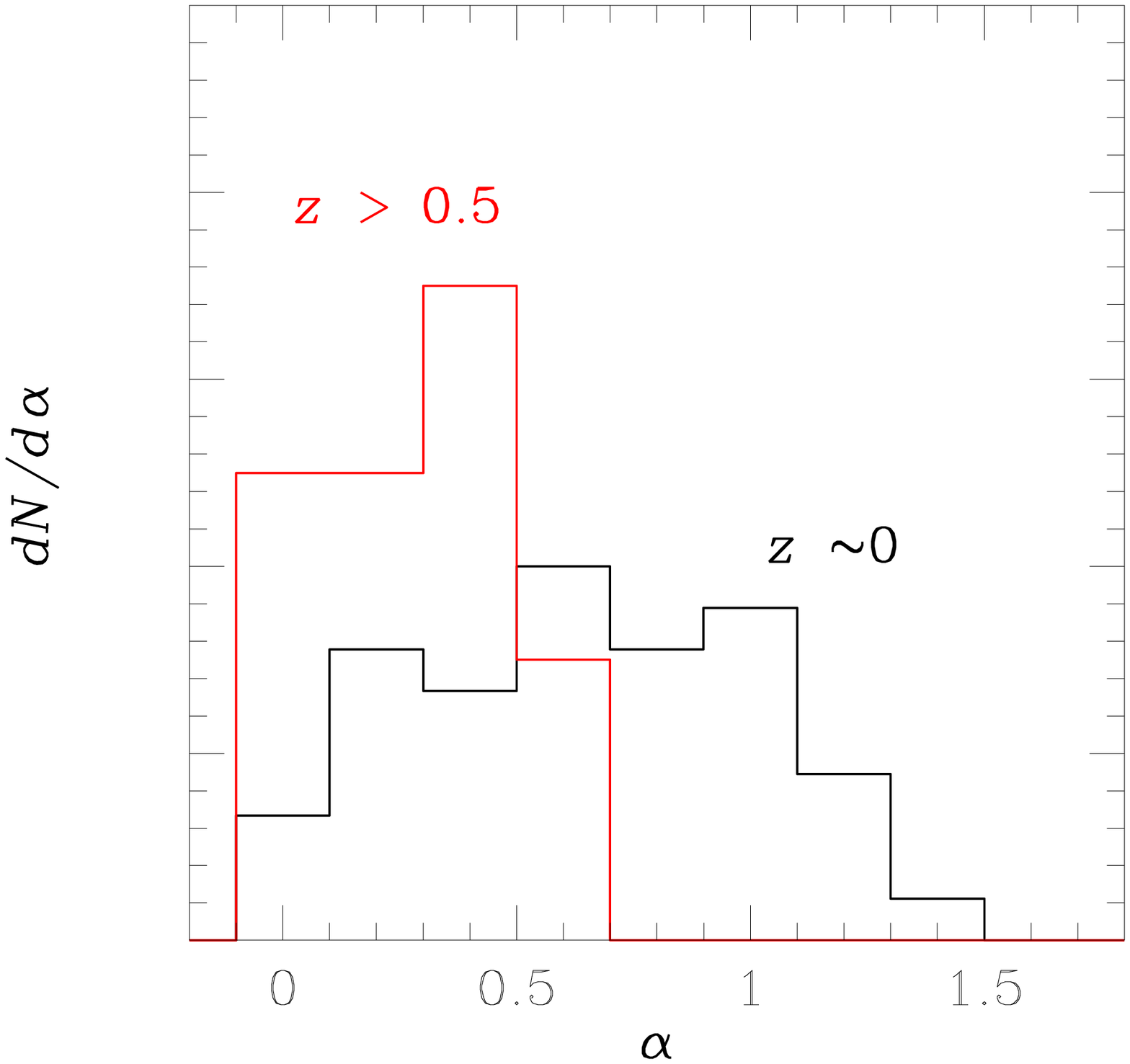}\hspace*{-1.5em}}
\caption{\emph{(a)} \emph{Chandra} images of the 400d clusters with
  $z>0.5$. Note a high fraction of objects that show clear signs of a
  major merger. \emph{(b)} Distribution of the cuspiness parameter for
  the $z>0.5$ and $z\sim0$ samples.} 
\label{fig:structure}
\end{figure}

\emph{Chandra} images of the $z>0.5$ objects from the 400d sample show a
clear evolution of the cluster X-ray morphologies --- at least 15 of 20
objects shows signs of an on-going major merger
(Fig.\,\ref{fig:structure}a); the corresponding fraction in the
low-redshift sample is $\lesssim 30\%$. The same effect is apparent in
the distribution of the cuspiness parameter shown in
Fig.\,\ref{fig:structure}b. Only 3 of 20 high-$z$ clusters have
$\alpha>0.5$ (i.e., above the boundary between cooling flow and non
cooling clusters, see \S~\ref{sec:oper-defin-cool}), while in the
low-$z$ sample this fraction is 31 of 48 (65\%). The are no clusters
with $\alpha>0.7$ (strong cooling flows) in the $z>0.5$ sample, while
the fraction of such clusters at $z\sim0$ is 46\% (22 of 48 objects).
The statistical significance of the difference in the distribution
corresponds to a random fluctuation probability of only
$P\simeq5\times10^{-6}$.

\bigskip

\noindent
Our results provide a tantalizing evidence for a strong evolution in the
incidence rate of the cluster cooling flows at $z>0.5$. This is
apparently related to the higher cluster merging rate, indeed expected
at these redshifts in the Dark Energy dominated, cold dark matter
cosmological models (e.g.,~\cite{2001ApJ...546..223G}). The cluster
cooling flows thus appear to be a relatively recent phenomenon, which
becomes common only in the past $1/3$ of the Hubble time.

%
%
% BibTeX users please use
\bibliographystyle{unsrt}
\bibliography{vikhlinin}
%%%%%%%%%%%%%%%%%%%%%%%%%%%%%%%%%%%%%%%%%%%%%%%%%%%%%%%%%%%%%%%%%%%%%%  }

%%%%%%%%%%%%%%%%%%%%%%%%%%%%%%%%%%%%%%%%%%%%%%%%%%%%%%%%%%%%%%%%%%%%%%

\printindex
\end{document}